\documentclass[preprint, aps]{revtex4}

\usepackage[english]{babel}
\usepackage{xcolor, dsfont, bbm}
\usepackage{pgf}
\usepackage{comment}
\usepackage{times}
\usepackage{graphics}
\usepackage{wrapfig}
\usepackage{psfrag, etoolbox}
\usepackage{amssymb, amsthm, amsmath, mathrsfs, appendix, bm}
\usepackage{lettrine}

\newtheorem{theorem}{Theorem}

\newtheorem{lemma}{Lemma}

\newcommand{\abs}[1]{\left\vert#1\right\vert}
\newcommand{\ass}{\stackrel{\textup{\tiny def}}{=}}

\newcommand{\cv}[1]{\Delta_{#1}}
\newcommand{\cvv}{\mathbf{C}}
\newcommand{\cvm}[1]{\Lambda_{#1}}
\newcommand{\ob}[1]{\hat{#1}}
\newcommand{\ex}[1]{E_{#1}}
\newcommand{\pp}{\tau}

\begin{document}



\title{Relativistic independence bounds nonlocality}

\author{Avishy Carmi}
\affiliation{Faculty of Engineering and the Center for Quantum Information Science and Technology, Ben-Gurion University of
  the Negev, Beersheba 8410501, Israel}
\email{avcarmi@bgu.ac.il}

\author{Eliahu Cohen}
\affiliation{Faculty of Engineering and the Institute of Nanotechnology and Advanced
Materials, Bar Ilan University, Ramat Gan 5290002, Israel}
\email{eliahu.cohen@biu.ac.il}

%
%



\begin{abstract}
If Nature allowed nonlocal correlations other than those predicted by
quantum mechanics, would that contradict some physical principle?
Various approaches have been put forward in the past two
decades in an attempt to single out quantum nonlocality. However, none
of them can explain the set of quantum correlations arising in the
simplest scenarios. Here it is shown that {\it generalized uncertainty relations}, as well as a specific notion of locality give rise to both familiar and
new characterizations of quantum correlations. In particular, we
identify a condition, {\it relativistic independence}, which states that
uncertainty relations are local in the sense that they cannot be
influenced by other experimenters' choices of measuring
instruments. We prove that theories with nonlocal correlations
stronger than the quantum ones do not satisfy this notion of locality
and therefore they either violate the underlying generalized uncertainty relations or allow
experimenters to nonlocally tamper with the uncertainty relations of
their peers.

\end{abstract}

\maketitle

\section{Introduction}
Quantum mechanics stands out in enabling strong, nonlocal correlations
between remote parties. On the one hand, these quantum correlations cannot in
any way be explained by models of classical physics. On
the other hand, quantum theory remains rather elusive
about their physical origin~\cite{PR1,SuperQ, Bellnonlocality:14}. What if Nature allowed nonlocal
correlations other than those predicted by quantum mechanics -- would
that break any known physical principle?  This question becomes all
more important when the predictions of quantum mechanics are
experimentally verified time and again.

Initially it was speculated that those correlations excluded by quantum
mechanics violate relativistic causality -- \emph{the principle
which dictates that experiments can be influenced only by events in
their past light cone, and influence events only in
their future light cone}. But then it was shown that other theories may exist
whose correlations, while not realizable in quantum mechanics, are
nevertheless non-signaling and are hence consistent with relativistic
causality~\cite{PR1}.

Over the past 20 years, many efforts have
been invested in a line of research aimed at quantitatively deriving
the strength of quantum correlations from basic principles.
For example, it was shown that violations of the Bell--CHSH
inequality~\cite{CHSH} beyond the quantum limit, known
  as Tsirelson's bound, are inconsistent with the uncertainty
principle \cite{unc2}. Popescu-Rohrlich--boxes
(PR--boxes), the hypothetical models achieving the maximal violation
of the Bell--CHSH inequality~\cite{PR1}, would
allow distributed computation to be performed with only one bit of
communication \cite{vanDam}, which looks unlikely but does not violate
any known physical law. Similarly, in stronger-than-quantum nonlocal
theories some computations exceed reasonable performance
limits~\cite{nonlocalcomp}, and there is no sensible
measure of mutual information between pairs of
systems~\cite{Infocause}. Finally, it was shown that
superquantum nonlocality does not permit classical physics to emerge
in the limit of infinitely many microscopic
systems~\cite{ML, Gisin,Rohrlich}, and also violates the exclusiveness
  of local measurement outcomes in multipartite settings
  \cite{LO}. However, none of these and other principles that
have been proposed~\cite{SuperQ} can explain the set
of one- and two-point correlators that fully characterize
the quantum probability distributions witnessed in the simplest bipartite
two-outcome scenario.


A consequence of relativistic causality within the framework
of probabilistic theories is known as the no-signaling condition --
the local probability distributions of one experimenter (marginal
probabilities) are independent of another experimenter's
choices~\cite{PR1}.
While the no-signaling condition is insufficient
to single out quantum correlations, it is shown here that an analogous requirement
applicable in conjunction with generalized uncertainty relations
is satisfied exclusively by quantum mechanical correlations.

\section{Results}

 In what follows we first assume (Subsection \ref{A})
  that {\it generalized uncertainty relations} are valid within the
  theory in question. Such uncertainty relations broaden the meaning
  of uncertainty beyond the realm of quantum mechanics, and give rise
  to the Schr\"{o}dinger-Robertson uncertainty relation when applied
  to the latter. Then in Subsection \ref{B}, we assume in addition a
  certain form of independence we name {\it relativistic
    independence}, meaning here that local uncertainty relations
  cannot be affected at a distance. The above assumptions accord well
  with experimental observations, yet generalize the underlying
  theoretical model beyond the quantum formalism.

\subsection{Generalized uncertainty relations} \label{A}

\begin{figure*}[htb]
\centering
\psfrag{v}[c]{\small RI: $r_{jk}=r$, \; \emph{independent of} $j,k$}
\psfrag{b}[c]{\small $_{\cv{A_1}^2}$}
\psfrag{y}[c]{\small $_{\cv{A_0}^2}$}
\psfrag{c}[c]{\small $_{\cv{B_j}^2}$}
\psfrag{x}[c]{\small $i$}
\psfrag{n}[c]{\small $k$}
\psfrag{z}[c]{\small $_{\cv{C_k}^2}$}
\psfrag{t}[c]{\small $\varrho^{AC}_{ik}$}
\psfrag{r}[c]{\small $\varrho^{BC}_{jk}$}
\psfrag{s}[c]{\small $\varrho^{AB}_{ij}$}
\psfrag{m}[c]{\small $j$}
\psfrag{g}[c]{\small $_{r_{jk}}$}
\psfrag{f}[c]{\small $_{\varrho^{BC}_{jk}}$}
\psfrag{h}[c]{\small $_{\varrho^{AB}_{1j}}$}
\psfrag{k}[c]{\small $_{\varrho^{AC}_{1k}}$}
\psfrag{q}[c]{\small $_{\varrho^{AB}_{0j}}$}
\psfrag{a}[c]{\small $_{\varrho^{AC}_{0k}}$}
\psfrag{p}[l]{$\succeq 0$}
\psfrag{d}[c]{$_{\emph{Alice}}$}
\psfrag{e}[c]{$_{\emph{Bob}}$}
\psfrag{j}[c]{\; \; $_{\emph{Charlie}}$}

\includegraphics[width=0.92\textwidth]{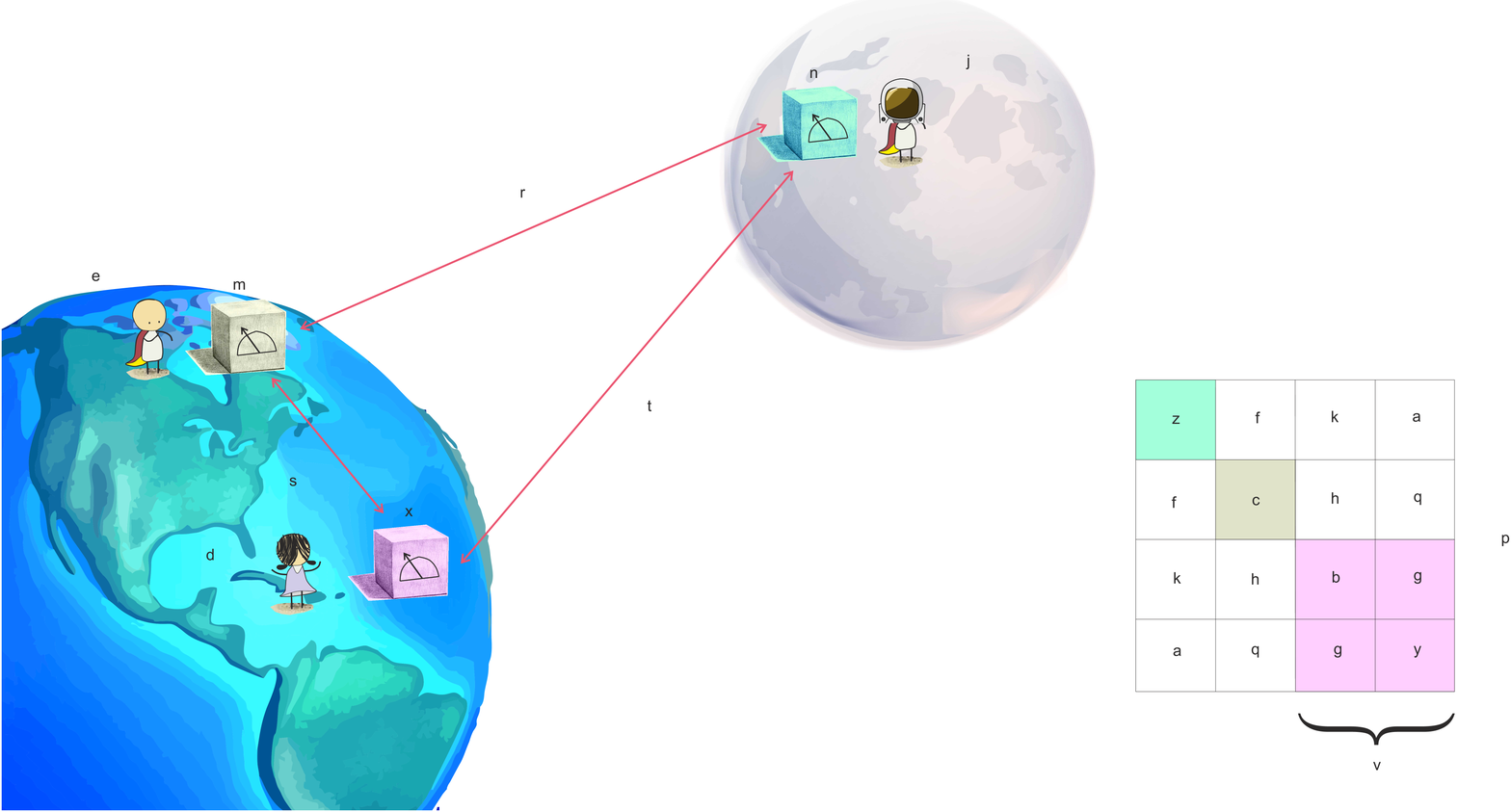}

\caption{\small An illustration of relativistic independence in a tripartite scenario. In a theory obeying generalized uncertainty relations (shown in the bottom right corner in the form of a certain positive-semidefinite matrix), relativistic independence (RI) prevents Bob and Charlie from influencing Alice's uncertainty relations, e.g., $\cv{A_0}^2 \cv{A_1}^2 \geq r_{jk}^2$, through their choices $j$ and $k$, i.e. $r_{jk}=r$. Here, $\varrho^{AB}_{ij} =
  \cvv(A_i, B_j)$, $\varrho^{AC}_{ik} =
  \cvv(A_i, C_k)$, and $\varrho^{BC}_{jk} =
  \cvv(B_j, C_k)$, illustrated by the arrows are the covariances of Alice-Bob, Alice-Charlie,
  and Bob-Charlie measurements, respectively.
  In the quantum mechanical formalism a similar matrix
  inequality gives rise to the Schr\"{o}dinger-Robertson uncertainty
  relations of Alice's self-adjoint operators $\ob{A}_0$ and
  $\ob{A}_1$, as well as between the nonlocal Alice-Bob operators,
  $\ob{A}_0 \ob{B}_j$ and $\ob{A}_1 \ob{B}_j$. See Materials and Methods.}
\label{fig:kid}
\end{figure*}

Three experimenters, Alice, Bob, and Charlie, perform an
experiment, where each of them owns a measuring
device. On each such device a knob determines its mode of
operation, either ``0'' or ``1'', which allows measuring two physical
variables, $A_0$/$A_1$ on Alice's side, $B_0$/$B_1$ on Bob's side, and
$C_0$/$C_1$ on Charlie's side. Alice and Bob are close to one another
and so they use the readings from all their devices to empirically
evaluate the variances, $\cv{A_i}^2$, $\cv{B_j}^2$, and the
covariances, $\cvv(A_i, B_j) \ass \ex{A_i B_j} - \ex{A_i}
\ex{B_j}$, where $\ex{A_i}$, $\ex{B_j}$, and $\ex{A_i B_j}$ are the
    respective \emph{one- and two-point
        correlators}. Charlie, on the other hand, is far from
    them. See Figure~\ref{fig:kid}.

Assume that measurements of physical variables are generally inflicted with
uncertainty. Not only does this uncertainty affect pairs of local measurements
performed by individual experimenters, it also governs any number of
measurements performed by groups of remote experimenters. In our
tripartite setting, for example, the measurements of Alice, Bob, and
Charlie are assumed to be jointly governed by the generalized uncertainty relation,
\begin{equation}
\label{eq:u}
\small
\cvm{ABC} \ass
\begin{bmatrix}
\cvm{C} & \cvv(B, \, C)^T & \cvv(A, \, C)^T \\
\cvv(B, \, C) & \cvm{B} & \cvv(A, \, B)^T \\
\cvv(A, \, C) & \cvv(A, \, B) & \cvm{A}
\end{bmatrix} \succeq 0
\end{equation}
which means that $\cvm{ABC}$ is a positive semidefinite matrix.
Here, $\cvv(A, \, B)$, $\cvv(A, \, C)$, and $\cvv(B, \, C)$ are the
empirical covariance matrices of Alice-Bob, Alice-Charlie, and Bob-Charlie
measurements. The diagonal submatrices, e.g., $\cvm{A}$, represent the
uncertainty relations governing the individual experimenters.
Below and in Materials and Methods, \eqref{eq:u} is shown to imply the quantum
mechanical Schr\"{o}dinger-Robertson uncertainty relations~\cite{Schr}, as well as
their multipartite non-quantum generalizations. Moreover, in local
hidden variables theories where all measurement outcomes preexist,
\eqref{eq:u} coincides with a covariance matrix, which is by
construction positive semidefinite and represents the
uncertainty of $A_i$, $B_j$, and $C_k$, hence the natural
generalization to other theories.

Provided that Bob measured $B_j$ and Charlie measured $C_k$, the
system as a whole is governed by a submatrix of $\cvm{ABC}$,
\begin{equation}
\label{eq:u3}
\small
\cvm{ABC}^{jk} \ass
\begin{bmatrix}
\cv{C_k}^2 & \cvv(C_k, \, B_j) & \cvv(C_k, \, A_1) & \cvv(C_k, \, A_0) \\
\cvv(C_k, \, B_j) & \cv{B_j}^2 & \cvv(B_j, \, A_1) & \cvv(B_j, \, A_0) \\
\cvv(C_k, \, A_1) & \cvv(B_j, \, A_1) & \cv{A_1}^2 & r_{jk} \\
\cvv(C_k, \, A_0) & \cvv(B_j, \, A_0) & r_{jk} & \cv{A_0}^2
\end{bmatrix} \succeq 0
\end{equation}
Here, $r_{jk}$ is
a real number whose value guarantees that $\cvm{ABC}^{jk} \succeq
0$. Therefore, it generally depends not only on Alice's choices
but also on Bob's $j$ and Charlie's $k$.
The lower $2 \times 2$ submatrix in \eqref{eq:u3}, which is henceforth denoted
as the positive-semidefinite $\cvm{A}^{jk}$, implies that Alice's measurements satisfy $\cv{A_0}^2
\cv{A_1}^2 \geq r_{jk}^2$, as well as other uncertainty relations that
depend on $r_{jk}$ rather than $r_{jk}^2$, i.e., $u^T \cvm{A}^{jk} u \geq 0$,
where $u$ is any two-dimensional real-valued vector.

Local hidden variables theories, quantum mechanics, and
non-quantum theories such as the hypothetical PR--boxes~\cite{PR1}
obey \eqref{eq:u3}. Moreover, they provide different closed forms for this $r_{jk}$,
which in general we are unable to assume. In local hidden variables theories, where $A_0$ and
$A_1$ are classical random variables whose joint probability
distribution is well-defined, \eqref{eq:u3} holds for $r_{jk} =
\cvv(A_0,\, A_1)$, which is independent of $j$ and $k$.
In quantum mechanics the Schr\"{o}dinger-Robertson uncertainty
relations show that $r_{jk}$ depends exclusively on Alice's
self-adjoint operators, in particular their commutator and
anti-commutator. If Alice and Charlie share a PR--box then $r_{jk} =
(-1)^k$, which, in contrast to the two other theories, depends on
$k$.\\[1ex]

\subsection{Independence} \label{B}

In the above setting, Bob and Charlie may be able to nonlocally tamper with
Alice's uncertainty relation, $\cvm{A}^{jk} \succeq 0$, through their
$j$ and $k$. Prohibiting this by requiring that Alice's uncertainty
relation as a whole, i.e., the trio $\cv{A_0}$, $\cv{A_1}$, and
$r_{jk}$, would be independent of Bob's $j$ and Charlie's $k$ leads to
the set of quantum mechanical one- and two-point correlators. This
condition is named henceforth \emph{relativistic independence} (RI).
%
%

By RI, the Alice-Bob system, which is governed by
the lower $3 \times 3$ submatrix of $\cv{ABC}^{jk}$, satisfies
$\cvm{A}^{jk} \ass \cvm{A}$, for $r_{jk} \ass r$. Swapping the roles of Alice and Bob,
where Alice measures $A_i$, RI similarly implies
$\cvm{B}^{ik} \ass \cvm{B}$, for $\bar{r}_{ik} \ass
\bar{r}$. In other words, RI means
\begin{equation}
\label{eq:uncg}
\small
\begin{array}{l}
\begin{bmatrix}
\cv{B_j}^2 & \cvv(B_j, \, A_1) & \cvv(B_j, \, A_0) \\
\cvv(B_j, \, A_1) & \cv{A_1}^2 & r \\
\cvv(B_j, \, A_0) & r & \cv{A_0}^2
\end{bmatrix} \succeq 0, \\
 \\
\begin{bmatrix}
\cv{A_i}^2 & \cvv(A_i, \, B_1) & \cvv(A_i, \, B_0) \\
\cvv(A_i, \, B_1) & \cv{B_1}^2 & \bar{r} \\
\cvv(A_i, \, B_0) & \bar{r} & \cv{B_0}^2
\end{bmatrix} \succeq 0,
\end{array}
\end{equation}
for $i,j \in \{0,1\}$.
%
RI \eqref{eq:uncg} and no-signaling are distinct and do not follow
from one another. The no-signaling condition, for example, dictates that the
(marginal) probability distributions of Alice's measurements, and
therefore also $\cv{A_0}^2$ and $\cv{A_1}^2$, are independent of Bob's
choices. RI, on the other hand, implies that $\cvm{A}$ in its entirety
must be independent of Bob's choices, which may hold whether
or not Alice's marginal probabilities are independent of $j$. The
relationship between the two conditions is discussed in more detail in
the Materials and Methods section.


PR--boxes satisfy the no-signaling condition but violate RI (see Materials and Methods). Moreover, as stated below, RI \eqref{eq:uncg} is satisfied
exclusively by the quantum mechanical bipartite one- and two-point correlators.

\begin{theorem}
\label{thm:main}
The conditions \eqref{eq:uncg} imply
\begin{equation}
\label{eq:qb}
\begin{array}{l}
\abs{\varrho_{00} \varrho_{10} - \varrho_{01} \varrho_{11} } \leq
\sum_{j=0,1} \sqrt{(1 - \varrho_{0j}^2) (1- \varrho_{1j}^2)} \\[1ex]
\abs{\varrho_{00} \varrho_{01} - \varrho_{10} \varrho_{11} } \leq
\sum_{i=0,1} \sqrt{(1 - \varrho_{i0}^2) (1- \varrho_{i1}^2)}
\end{array}
\end{equation}
where $\varrho_{ij} \ass \cvv(A_i, \, B_j) /
(\cv{A_i} \cv{B_j})$, is the Pearson correlation
  coefficient between $A_i$ and $B_j$.
\end{theorem}

It is known that any four correlators, $\ex{A_i B_j}$, must
satisfy \eqref{eq:qb} if they are to describe the nonlocality present
in a physically realizable quantum mechanical pair of
systems~\cite{Bellnonlocality:14}. In addition, all the sets of such
correlators permitted by \eqref{eq:qb} are possible within quantum
mechanics. This result was proven when assuming quantum mechanics and
vanishing one-point correlators, $\ex{A_i} = \ex{B_j} = 0$,
independently by Tsirelson, Landau, and
Massanes~\cite{T87,L88,M03}. More recently, \eqref{eq:qb} has been
derived for the case of binary measurements from the first level of the NPA
hierarchy~\cite{NPA}. We show without assuming any
of these that this bound (in the form of Landau) originates from RI \eqref{eq:uncg}.
Moreover,
it is now clear that \eqref{eq:qb} must hold not only for binary but also for other, both
discrete and continuous, variables. Consequently, Tsirelson's
$2 \sqrt{2}$ bound~\cite{T80} on the Bell-CHSH parameter~\cite{CHSH},
$\mathscr{B}_{AB} \ass \varrho_{00} + \varrho_{10} +
  \varrho_{01} - \varrho_{11}$, applies to any type of measurement. For example,
Alice's and Bob's measurements may be the position and momentum of some
wavefunction. 

Quantum theory satisfies the RI condition \eqref{eq:uncg}
and is therefore subject to \eqref{eq:qb}. Furthermore, in the
case of binary $\pm 1$ measurements whose one-point correlators vanish, the first
Alice-Bob uncertainty relation in \eqref{eq:uncg} is given in quantum
mechanics by the Schr\"{o}dinger-Robertson uncertainty relations of $\ob{A}_0 \ob{B}_j$
and $\ob{A}_1 \ob{B}_j$, where $\ob{A}_i$ and $\ob{B}_j$ are Alice's and
Bob's self-adjoint operators. See Materials and Methods for the proof of this theorem and
for further details.

Surprisingly, within the quantum formalism \eqref{eq:qb} is a special
case of another bound with two extra terms.

\begin{theorem}
\label{thm:main1}
In quantum theory, where the Alice and Bob
measurements are represented by the self-adjoint operators $\ob{A}_i$ and
$\ob{B}_j$, the following holds, 
\begin{equation}
\label{eq:qbq}
\begin{array}{l}
\abs{\varrho_{00} \varrho_{10} - \varrho_{01} \varrho_{11} } \leq
\sum_{j=0,1} \sqrt{(1 - \varrho_{0j}^2) (1- \varrho_{1j}^2) - \eta_{\ob{A}}^2} \\[1ex]
\abs{\varrho_{00} \varrho_{01} - \varrho_{10} \varrho_{11} } \leq
\sum_{i=0,1} \sqrt{(1 - \varrho_{i0}^2) (1- \varrho_{i1}^2) - \eta_{\ob{B}}^2}
\end{array}
\end{equation}
where $\varrho_{ij} \ass \left(\langle \ob{A}_i \ob{B}_j \rangle -
  \langle \ob{A}_i \rangle \langle \ob{B}_j
  \rangle \right) / \left( \cv{\ob{A}_i} \cv{\ob{B}_j} \right)$, and
$\eta_{\ob{X}} \ass \frac{1}{2i} \langle [\ob{X}_0, \ob{X}_1]
  \rangle / \left( \cv{\ob{X}_0} \cv{\ob{X}_1} \right)$, with $\ob{X}$ being either
$\ob{A}$ or $\ob{B}$. Here, $[\ob{X}_0,\ob{X}_1] \ass \ob{X}_0 \ob{X}_1
- \ob{X}_1 \ob{X}_0$ is the commutator of $\ob{X}_0$ and $\ob{X}_1$,
and $\cv{\ob{X}}^2 = \langle \ob{X}^2 \rangle - \langle \ob{X}
\rangle^2$ is the variance of $\ob{X}$. The $\langle \cdot \rangle$ is
the quantum-mechanical expectation. Note that $\frac{1}{2i}
[\ob{X}_0,\ob{X}_1]$ is self-adjoint and is therefore an
observable. Moreover, $\abs{\eta_{\ob{X}}} \leq 1$, where
$\abs{\eta_{\ob{X}}} = 1$ only if the Robertson uncertainty relation of
$\ob{X}_0$ and $\ob{X}_1$ is saturated.
\end{theorem}
\noindent The proof of this theorem is given in Materials and Methods.


\begin{figure*}[htb]
\centering
\psfrag{a}[c]{\small $\varrho_{00} \varrho_{10}$}
\psfrag{b}[c]{\small $\varrho_{01} \varrho_{11}$}
\psfrag{c}[c]{\small $\sigma_{00} \sigma_{10}$}
\psfrag{d}[c]{\small $\sigma_{01} \sigma_{11}$}
\psfrag{1}[c]{\small $_{\eta_{\ob{A}}}$}
\psfrag{0}[l]{\small $_{\nu_{\ob{A}}}$}
\psfrag{8}[r]{\small $\frac{\hbar / 2}{\cv{\hat{x}} \cv{\hat{p}}}$}
\psfrag{9}[l]{\small $\frac{\mathscr{B}}{2 \sqrt{2}}$}
\psfrag{s}[c]{\small $\left(\frac{\mathscr{B}}{2 \sqrt{2}}\right)^2 +
  \left( \frac{\hbar / 2}{\cv{\hat{x}} \cv{\hat{p}}} \right)^2 \leq 1$}
\psfrag{m}[c]{$\emph{the Tsirelson bound}$}

\includegraphics[width=0.80\textwidth]{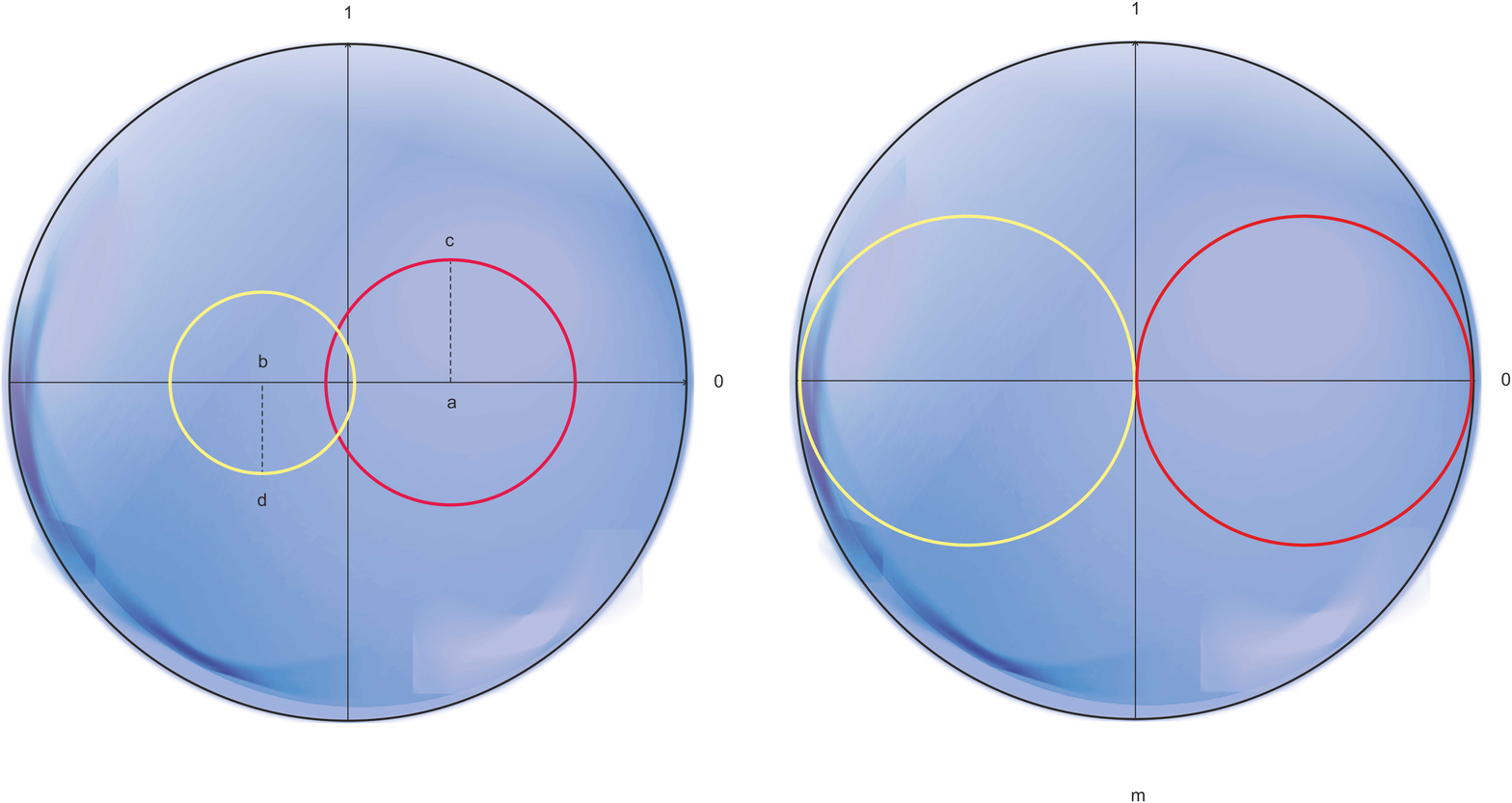}
\caption{\small Geometry of bipartite relativistic independence in Hilbert
  space, the bounds \eqref{eq:qbq}. The $\eta_{\ob{A}}$ is as defined in Theorem~\ref{thm:main1},
  and $\nu_{\ob{A}} \ass \left (\frac{1}{2} \langle \{\ob{A}_0, \ob{A}_1 \} \rangle - \langle \ob{A}_0 \rangle
  \langle \ob{A}_1 \rangle \right) / \left( \cv{\ob{A}_0}
  \cv{\ob{A}_1}\right)$, where $\{\ob{X}, \ob{Y}\}$ is the
anti-commutator. Using these definitions the Schr\"{o}dinger-Robertson
uncertainty relation between Alice's observables is $\nu_{\ob{A}}^2 + \eta_{\ob{A}}^2
\leq 1$, hence the pair of bluish unit disks. Bob's choice, $j=1$ or $j=0$, further confines
Alice's uncertainty, the $\eta_{\ob{A}}$ and $\nu_{\ob{A}}$, to one of
the circles, the yellow or the red, respectively. The extent and location of these circles is determined by the
nonlocal covariances, $\varrho_{ij}$. Quantum mechanics satisfies relativistic
independence and thus keeps Alice's uncertainty relations independent
of Bob's choices, i.e., by allowing only those covariances for which
the red and yellow circles intersect. Tsirelson's bound is an extreme
configuration where these circles intersect at the origin.}

\label{fig:ind}
\end{figure*}

\subsection{Local uncertainty relations and nonlocal correlations}

The geometry of bipartite RI in Hilbert space is illustrated in
Figure~\ref{fig:ind}.
The left picture in this figure is the geometry
underlying the first bound in \eqref{eq:qbq}. This bound arises from the two
  uncertainty relations \eqref{eq:uncg}, which from within quantum mechanics
  coincide with the Schr\"{o}dinger-Robertson uncertainty relations of
  $\ob{A}_0 \ob{B}_j$ and $\ob{A}_1 \ob{B}_j$ in
  the special case of binary measurements. In other cases,
  \eqref{eq:uncg} may be viewed as a generalization of the
  Schr\"{o}dinger-Robertson uncertainty relations. As shown in Materials and Methods,
  inside Hilbert space \eqref{eq:uncg} describe two circles in the
  complex plane, one for $j=0$ (red) and another for $j=1$ (yellow).
  The circles are centered at $\varrho_{0j} \varrho_{1j}$, and their respective radiuses are
$\sigma_{0j} \sigma_{1j}$, where $\sigma_{ij}^2 = 1 -
\varrho_{ij}^2$. Alice's local uncertainty relations are
  confined to one or another circle depending on Bob's choice $j$. Quantum mechanics satisfies RI and thus
keeps Alice's uncertainty relations independent of Bob's choice,
i.e., by allowing only those covariances $\varrho_{ij}$ for which
the red and yellow circles intersect. Tsirelson's bound (the right picture), for
example, is attained when the region of intersection collapses to a
single point at the origin.

RI implies that the extent of nonlocality is governed by local
uncertainty relations. The interplay between nonlocality as
quantified by the Bell-CHSH parameter, $\mathscr{B}$, and Heisenberg
uncertainty where $\ob{A}_0 = \hat{x}$ and $\ob{A}_1=\hat{p}$, are the
position and momentum operators (See  Materials and Methods for the complete
derivation), is
\begin{equation}
\left( \frac{\mathscr{B}}{2 \sqrt{2}} \right)^2 + \left( \frac{\hbar /
      2}{\cv{\hat{x}} \cv{\hat{p}} } \right)^2 \leq 1.
\end{equation}

It is known that a complete characterization of the set of quantum
correlations must follow from inherently multipartite principles~\cite{multi}.
Indeed, as shown in the Materials and Methods section, RI applies to any number of
parties with any number of measuring devices. This allows, for
example, deriving a generalization of \eqref{eq:qb} for the Alice-Bob,
Alice-Charlie, and Bob-Charlie one- and two-point correlators in a
tripartite scenario. The property known as monogamy of correlations, the
$\abs{\mathscr{B}_{AB}} + \abs{\mathscr{B}_{AC}} \leq 4$, follows as a
special case of this inequality. In the same section,
it is shown that the correlators in local hidden variable theories can be
similarly bounded by a variant of RI.

\section{Discussion}
Within a class of theories obeying generalized uncertainty relations,
relativistic independence was shown to reproduce the complete quantum mechanical
characterization of the bipartite correlations in two-outcome
scenarios, and potentially in much more general cases as
straightforward corollaries of our approach. To fully characterize the
set of quantum correlations would generally require analyzing the
uncertainty relation \eqref{eq:u} in an elaborate multipartite setting, accounting
for all the parties' cross-correlations and assuming
relativistic independence (this point, as well as some other technical issues, are discussed in detail within the Materials and Methods section). All these imply that
stronger-than-quantum nonlocal theories may either be incompatible with
the uncertainty relations analyzed above or allow experimenters to
nonlocally tamper with the uncertainty relations of other experimenters.\\[2ex]

\section{Materials and Methods}

\subsection{No-signaling and relativistic independence}

A consequence of relativistic causality in probabilistic theories is
the no-signaling condition~\cite{PR1}. Consider the Bell-CHSH setting
where $a$ and $b$ are the outcomes of Alice's and Bob's measurements.
The joint probability of these outcomes when Alice measured using
device $i$ and Bob using device $j$ is denoted as $p(a,b \mid
i,j)$. No-signaling states that one experimenter's marginal
probabilities are independent of another experimenter's choices,
namely,
\begin{equation}
\begin{array}{l}
\sum_b p(a, b \mid i, \, 0) = \sum_b p(a, b \mid i, \, 1)  \ass p(a
\mid i) \\[1ex]
\sum_a p(a, b \mid 0, \, j) = \sum_a p(a, b \mid 1, \, j)  \ass p(b
\mid j)
\end{array}
\end{equation}
Of course it means that one experimenter's precision
is independent of another experimenter's choices,
\begin{equation}
\begin{array}{l}
\cv{A_i}^2 = E_{a^2 \mid i,j} - E_{a \mid i,j}^2 = \sum_{a,b} a^2 p(a,b
  \mid i,j) - \left( \sum_{a,b} a p(a,b \mid i,j) \right)^2 = \sum_a
  a^2 p(a \mid i) - \left( \sum_a a p(a \mid i) \right)^2\\[1ex]
\cv{B_j}^2 = E_{b^2 \mid i,j} - E_{b \mid i,j}^2 = \sum_{a,b} b^2 p(a,b
  \mid i,j) - \left( \sum_{a,b} b p(a,b \mid i,j) \right)^2 = \sum_b
  b^2 p(b \mid j) - \left( \sum_b b p(b \mid j) \right)^2
\end{array}
\end{equation}
The no-signaling condition thus implies that the variances of one
experimenter in the Alice-Bob uncertainty relations \eqref{eq:uncg}
are independent of the other experimenter's choices.

Relativistic independence implies that one experimenter's uncertainty relation is
altogether independent of the other experimenter's choices, i.e.,
that $\cvm{A}$ as a whole, and therefore also $r_j$, are independent
of $j$. This does not necessarily imply the no-signaling condition as
there may exist, for example, marginal distributions $p(a \mid i,j)$ that depend on Bob's
$j$ whose variances, $\cv{A_i}^2$, are nevertheless independent of
this $j$. This shows that relativistic independence does not at all require us to assume
the no-signaling condition.


\subsection{Popescu--Rohrlich boxes violate relativistic independence}

Consider a tripartite setting where Bob and Charlie are uncorrelated,
$\cvv(B_j, \, C_k) = 0$, and Alice and Charlie share a
PR-box~\cite{PR1}. The PR-boxes define, $\ex{A_i C_k} = (-1)^{ik}$,
and $\ex{A_i} = 0$, $\ex{C_k} = 0$. The variances
are thus, $\cv{A_i}^2 = \ex{A_i^2} - \ex{A_i}^2 = 1$ and $\cv{C_k}^2 = \ex{C_k^2} -
\ex{C_k}^2 = 1$, and the covariances are $\cvv(A_1, \, C_k) =
(-1)^k$ and $\cvv(A_0, \, C_k) = 1$. In this case, a permutation of
\eqref{eq:u3} reads
\begin{equation}
\small
\cvm{PR}^{jk} \ass
\begin{bmatrix}
\cv{B_j}^2 & 0 & \cvv(A_1, \, B_j) & \cvv(A_0, \, B_j) \\
0 & 1 & 1 & (-1)^k \\
\cvv(A_1, \, B_j) & 1 & 1 & r_{jk} \\
\cvv(A_0, \, B_j) & (-1)^k & r_{jk} & 1
\end{bmatrix}
\succeq 0
\end{equation}
Namely,
\begin{equation}
\label{eq:pr1}
\small
M^{-1} \cvm{PR}^{jk} M^{-1} =
\begin{bmatrix}
1 & 0 & \varrho^{AB}_{1j} & \varrho^{AB}_{0j} \\
0 & 1 & 1 & (-1)^k \\
\varrho^{AB}_{1j} & 1 & 1 & r_{jk} \\
\varrho^{AB}_{0j} & (-1)^k & r_{jk} & 1
\end{bmatrix}
\succeq 0
\end{equation}
where $M$ is a diagonal matrix whose (non-vanishing) terms are all
ones but $\cv{B_j}$. By the Schur complement condition for positive
semidefiniteness, \eqref{eq:pr1} is equivalent to
\begin{equation}
\label{eq:pr2}
\small
\begin{bmatrix}
1 & r_{jk} \\
r_{jk} & 1
\end{bmatrix}
\succeq
\begin{bmatrix}
\varrho^{AB}_{1j} & 1 \\
\varrho^{AB}_{0j} & (-1)^k
\end{bmatrix}
\begin{bmatrix}
\varrho^{AB}_{1j} & 1 \\
\varrho^{AB}_{0j} & (-1)^k
\end{bmatrix}^T
=
\begin{bmatrix}
\left(\varrho^{AB}_{1j}\right)^2 &  \varrho^{AB}_{1j} \varrho^{AB}_{0j}\\
\varrho^{AB}_{1j} \varrho^{AB}_{0j} & \left(\varrho^{AB}_{0j}\right)^2
\end{bmatrix} +
\begin{bmatrix}
1 & (-1)^k \\
(-1)^k & 1
\end{bmatrix}
\end{equation}
which renders $\varrho^{AB}_{ij} = 0$ (positive-semidefiniteness of the matrix obtained by subtracting the right-hand side from the left-hand side implies the non-negativity of its diagonal entries from which this result follows). The inequality \eqref{eq:pr2}
is equivalent to $-[r_{jk} - (-1)^k]^2 \geq 0$, and only holds for
$r_{jk} = (-1)^k$. Such a theory therefore violates
relativistic independence.

But the PR-box example teaches us something profound. In this model,
complementarity (i.e., the inability to measure both local variables
in the same experiment) must be assumed in both Alice's and Charlie's ends, for
otherwise Alice, for example, may evaluate,
\begin{equation}
\label{eq:gah}
A_0 A_1 = (A_0C_k)(A_1C_k) = \cvv(A_0, \, C_k)\cvv(A_1, \, C_k) = (-1)^0(-1)^k = (-1)^k=r_{jk}
\end{equation}
from which she could tell Charlie's choice $k$. Lack of complementarity
immediately leads to signaling in the case of PR-boxes, but as we have
seen, the weaker assumption of uncertainty leads to a problem
with relativistic independence.



\subsection{Schr\"{o}dinger-Robertson uncertainty relations and the generalized
  uncertainty relations \eqref{eq:u}, \eqref{eq:u3}, and
  \eqref{eq:uncg}}

Let $\ob{A}_i$ and $\ob{B}_j$ be self-adjoint operators with $\pm 1$
eigenvalues and $\langle \ob{A}_i \rangle = \langle \ob{B}_j \rangle =
0$, whose product, $\ob{A}_i \ob{B}_j$ is similarly self-adjoint. The
Schr\"{o}dinger-Robertson uncertainty relations of the corresponding products,
$\ob{A}_0 \ob{B}_j$ and $\ob{A}_1 \ob{B}_j$,
\begin{equation}
\label{eq:Sun}
\small
\cv{\ob{A}_0 \ob{B}_j}^2 \cv{\ob{A}_1 \ob{B}_j}^2 \geq \left(
  \frac{1}{2} \langle \{\ob{A}_0, \ob{A}_1\} \rangle - \cvv(\ob{A}_0, \ob{B}_j)
  \cvv(\ob{A}_1, \ob{B}_j) \right)^2 + \left( \frac{1}{2i} \langle [\ob{A}_0,
  \ob{A}_1] \rangle \right)^2
\end{equation}
where $\cvv(\ob{A}_i, \ob{B}_j) = \langle \ob{A}_i \ob{B}_j
\rangle$, and the variance, $\cv{\ob{A}_i \ob{B}_j}^2 = 1 -
\cvv(\ob{A}_i, \ob{B}_j)^2$, can alternatively be written as
\begin{equation}
\small
\begin{bmatrix}
1 & \langle \ob{A}_0 \ob{A}_1 \rangle \\
\langle \ob{A}_1 \ob{A}_0 \rangle & 1
\end{bmatrix} \succeq
\begin{bmatrix}
\cvv(\ob{A}_1, \ob{B}_j)^2 & \cvv(\ob{A}_1, \ob{B}_j) \cvv(\ob{A}_0, \ob{B}_j) \\
\cvv(\ob{A}_1, \ob{B}_j) \cvv(\ob{A}_0, \ob{B}_j) & \cvv(\ob{A}_0, \ob{B}_j)^2
\end{bmatrix}
\end{equation}
By the Schur complement condition for positive semidefiniteness this is
equivalent to
\begin{equation}
\small
\begin{bmatrix}
\cv{\ob{B}_j}^2 & \cvv(\ob{A}_1, \ob{B}_j) & \cvv(\ob{A}_0, \ob{B}_j) \\
\cvv(\ob{A}_1, \ob{B}_j) & \cv{\ob{A}_1}^2  & \langle \ob{A}_0 \ob{A}_1 \rangle \\
\cvv(\ob{A}_0, \ob{B}_j) & \langle \ob{A}_1 \ob{A}_0 \rangle & \cv{\ob{A}_0}^2
\end{bmatrix} \succeq 0
\end{equation}
because $\cv{\ob{B}_j}^2 = \langle \ob{B}_j^2 \rangle - \langle \ob{B}_j
\rangle^2 = 1$ and $\cv{\ob{A}_i}^2 = \langle \ob{A}_i^2 \rangle -
\langle \ob{A}_i \rangle^2 = 1$. This in turn implies
\begin{equation}
\label{eq:sg}
\small
\begin{bmatrix}
\cv{\ob{B}_j}^2 & \cvv(\ob{A}_1, \ob{B}_j) & \cvv(\ob{A}_0, \ob{B}_j) \\
\cvv(\ob{A}_1, \ob{B}_j) & \cv{\ob{A}_1}^2  & r \\
\cvv(\ob{A}_0, \ob{B}_j) & r & \cv{\ob{A}_0}^2
\end{bmatrix} \succeq 0
\end{equation}
with $r = \langle \{\ob{A}_0, \ob{A}_1\} \rangle / 2$. The
inequalities in \eqref{eq:uncg} generalize the uncertainty relation
\eqref{eq:sg} to arbitrary measurements. The inequalities \eqref{eq:u}
and \eqref{eq:u3} further extend \eqref{eq:sg} to include the
remaining measurements of Alice, Bob and Charlie.



\subsection{Proof of Theorem~\ref{thm:main}}


By the Schur complement condition for positive semidefiniteness the
first condition in \eqref{eq:uncg} is equivalent to
$\cvm{A} \succeq \cv{B_j}^{-2} \cvv(A, B_j) \cvv(A, B_j)^T$. This can be normalized,
\begin{equation}
\label{eq:unc2}
M^{-1} \cvm{A} M^{-1} =
\begin{bmatrix}
1 & r' \\
r' & 1
\end{bmatrix}
\succeq \begin{bmatrix}
\varrho_{1j}^2 & \varrho_{0j} \varrho_{1j} \\
\varrho_{0j} \varrho_{1j} & \varrho_{0j}^2
\end{bmatrix}
= \cv{B_j}^{-2} M^{-1}
\begin{bmatrix}
\cvv(A_1, \, B_j) \\
\cvv(A_0, \, B_1)
\end{bmatrix}
\begin{bmatrix}
\cvv(A_1, \, B_j) & \cvv(A_0, \, B_1)
\end{bmatrix}
 M^{-1}
\end{equation}
where, $r' \ass \frac{r}{\cv{A_1} \cv{A_0}}$, and
$M$ is a diagonal matrix whose (non-vanishing) entries are
$\cv{A_1}$, and $\cv{A_0}$. This condition is equivalent to
\begin{equation}
\label{eq:covineq}
\abs{r' - \varrho_{0j}
  \varrho_{1j}} \leq \sqrt{(1-\varrho_{0j}^2) (1-\varrho_{1j}^2)}
\end{equation}
which follows from the non-negative determinant of the matrix obtained by
subtracting the right-hand side from the left-hand side in \eqref{eq:unc2}.
This together with the triangle inequality yield
\begin{equation}
\abs{\varrho_{00} \varrho_{10} - r' + r' - \varrho_{01} \varrho_{11}}
\leq \abs{r' - \varrho_{00} \varrho_{10}} + \abs{r' - \varrho_{01}
  \varrho_{11}} \leq \sum_{j=0,1} \sqrt{(1 - \varrho_{0j}^2) (1-
  \varrho_{1j}^2)}
\end{equation}
The second inequality in \eqref{eq:qb} is similarly obtained by
swapping the roles of Alice and Bob, i.e., from the second relativistic independence
condition in \eqref{eq:uncg}.

\subsection{Proof of Theorem~\ref{thm:main1}}

In the Hilbert space formulation of quantum mechanics Alice's measurements
are represented by the self-adjoint operators $\ob{A}_0$ and
$\ob{A}_1$. Similarly, Bob's measurements are represented by the
self-adjoint operators $\ob{B}_j$. The Schr\"{o}dinger-Robertson uncertainty
relations of $\ob{A}_0$ and $\ob{A}_1$ is
\begin{equation}
\label{eq:SR1}
\cv{\ob{A}_0}^2 \cv{\ob{A}_1}^2 \geq \left(
  \frac{1}{2} \langle \{ \ob{A}_0, \ob{A}_1 \} \rangle -
  \langle \ob{A}_0 \rangle \langle \ob{A}_1 \rangle
\right)^2 + \left( \frac{1}{2 i} \langle
[\ob{A}_0,\ob{A}_1] \rangle \right)^2
\end{equation}
where $\cv{\ob{A}_i}^2 = \langle \ob{A}_i^2 \rangle - \langle \ob{A}_i \rangle^2$
is the variance of $\ob{A}_i$. This may alternatively be written as
\begin{equation}
\label{eq:a}
\cvm{\ob{A}} =
\begin{bmatrix}
\cv{\ob{A}_1}^2 & r_{\text{Q}}\\
r_{\text{Q}}^{*} & \cv{\ob{A}_0}^2
\end{bmatrix} \succeq 0
\end{equation}
where $r_{\text{Q}} \ass \langle \ob{A}_1 \ob{A}_0 \rangle - \langle
  \ob{A}_1 \rangle \langle \ob{A}_0 \rangle$ with $r_Q^{*}$ being its
  complex conjugate. It can be recognized that this leads to Alice's part in the generalized uncertainty
  relation \eqref{eq:u3} where $r_{jk} = (r_{\text{Q}} +
  r_{\text{Q}}^{*}) / 2$ is independent of $j$ and $k$.

We shall show that the relativistic independence condition, the first inequality in
\eqref{eq:uncg}, holds in Hilbert space. This condition tells that
\begin{equation}
\label{eq:uncdd1}
\cvm{\ob{A} \ob{B}} =
\begin{bmatrix}
\cv{\ob{B}_j}^2 &  \langle \ob{A}_1
\ob{B}_j \rangle - \langle \ob{A}_1 \rangle \langle
\ob{B}_j \rangle & \langle \ob{A}_0
\ob{B}_j \rangle - \langle \ob{A}_0 \rangle \langle
\ob{B}_j \rangle\\
 \langle \ob{A}_1
\ob{B}_j \rangle - \langle \ob{A}_1 \rangle \langle
\ob{B}_j \rangle & \cv{\ob{A}_1}^2 &
\langle \ob{A}_1 \ob{A}_0 \rangle - \langle
  \ob{A}_1 \rangle \langle \ob{A}_0 \rangle \\
 \langle \ob{A}_0
\ob{B}_j \rangle - \langle \ob{A}_0 \rangle \langle
\ob{B}_j \rangle & \langle \ob{A}_0 \ob{A}_1 \rangle - \langle
  \ob{A}_1 \rangle \langle \ob{A}_0 \rangle & \cv{\ob{A}_0}^2
\end{bmatrix}, \quad j=0,1
\end{equation}
where $\cv{\ob{B}_j}^2 = \langle \ob{B}_j^2 \rangle -
\langle \ob{B}_j \rangle^2$, is a positive semidefinite matrix.
Let $U^{*}=[u_1,u_2,u_3]$ be any $3 \times 1$ complex-valued vector, and
denote $| \phi \rangle$ the underlying state. Note that
\begin{equation}
U^{*} \cvm{\ob{A} \ob{B}} U = V^{*} V  \geq 0
\end{equation}
where
\begin{equation}
V \ass u_1 \left(
    \ob{B}_j - \langle \ob{B}_j \rangle\right) | \phi \rangle +
  u_2 \left( \ob{A}_1 -\langle \ob{A}_1 \rangle \right) | \phi \rangle
  + u_3 \left( \ob{A}_0 - \langle \ob{A}_0 \rangle
  \right) | \phi \rangle
\end{equation}
which shows that $\cvm{\ob{A} \ob{B}} \succeq 0$, and therefore
\eqref{eq:uncg} hold.


In what follows we show that $\cvm{\ob{A} \ob{B}} \succeq 0$ implies
the first bound in \eqref{eq:qbq}. Note that
\begin{equation}
\label{eq:nn}
M^{-1} \cvm{\ob{A} \ob{B}} M^{-1} =
\begin{bmatrix}
1 & \varrho_{1j} & \varrho_{0j} \\
\varrho_{1j} & 1 & \frac{r_{\text{Q}}}{\cv{\ob{A}_1} \cv{\ob{A}_0}} \\
\varrho_{0j} & \frac{r_{\text{Q}}^{*}}{\cv{\ob{A}_1} \cv{\ob{A}_0}} & 1
\end{bmatrix} \succeq 0, \quad j=0,1
\end{equation}
where $M$ is a diagonal matrix whose (non-vanishing) entries are
$\cv{\ob{B}_j}$, $\cv{\ob{A}_1}$, and
$\cv{\ob{A}_0}$. By the Schur completment condition for positive
semidefiniteness \eqref{eq:nn} is equivalent to
\begin{equation}
\begin{bmatrix}
1 - \varrho_{1j}^2 & \frac{r_{\text{Q}}}{\cv{\ob{A}_1}
    \cv{\ob{A}_0}} - \varrho_{1j} \varrho_{0j} \\
\frac{r_{\text{Q}}^{*}}{\cv{\ob{A}_1}
    \cv{\ob{A}_0}} - \varrho_{1j} \varrho_{0j} & 1 -
\varrho_{0j}^2
\end{bmatrix} \succeq 0, \quad j=0,1
\end{equation}
This in turn is equivalent to the requirement that the determinant of
this matrix is nonnegative, i.e., that
\begin{equation}
\label{eq:ns}
\left( 1 - \varrho_{1j}^2 \right) \left( 1 - \varrho_{0j}^2 \right)
\geq \left( \frac{\langle \{ \ob{A}_0,\ob{A}_1 \} \rangle/2 -
  \langle \ob{A}_0 \rangle \langle
  \ob{A}_1 \rangle}{\cv{\ob{A}_1} \cv{\ob{A}_0}} -
\varrho_{0j} \varrho_{1j}  \right)^2 + \left( \frac{1}{2i}
\frac{\langle [
  \ob{A}_0,\ob{A}_1 ] \rangle}{\cv{\ob{A}_1}
    \cv{\ob{A}_0}}\right)^2, \quad j=0,1
\end{equation}
namely,
\begin{equation}
\sqrt{\left( 1 - \varrho_{1j}^2 \right) \left( 1 - \varrho_{0j}^2
  \right) - \eta_{\ob{A}}^2}
\geq \abs{ \frac{\langle \{ \ob{A}_0,\ob{A}_1 \} \rangle/2 -
  \langle \ob{A}_0 \rangle \langle
  \ob{A}_1 \rangle}{\cv{\ob{A}_1} \cv{\ob{A}_0}} -
\varrho_{0j} \varrho_{1j}}, \quad j=0,1
\end{equation}
where $\eta_{\ob{A}}$ is as defined in the theorem. This
together with the triangle inequality implies the first bound in
the theorem,
\begin{equation}
\abs{\varrho_{00} \varrho_{10} - \varrho_{01} \varrho_{11}} \leq
\sum_{j=0,1} \abs{ \frac{\langle \{ \ob{A}_0,\ob{A}_1 \} \rangle/2 -
  \langle \ob{A}_0 \rangle \langle
  \ob{A}_1 \rangle}{\cv{\ob{A}_1} \cv{\ob{A}_0}} -
\varrho_{0j} \varrho_{1j}} \leq \sum_{j=0,1} \sqrt{\left( 1 - \varrho_{1j}^2 \right) \left( 1 - \varrho_{0j}^2
  \right) - \eta_{\ob{A}}^2}
\end{equation}
The remaining bound similarly follows from the second relativistic independence
condition in \eqref{eq:uncg}. 

It is was previously noted that for the case where $\ob{A}_i^2 =
\ob{B}_j^2 = I$ and $\langle A_i \rangle = \langle B_j \rangle = 0$,
the inequality \eqref{eq:ns} coincides with the
Schr\"{o}dinger-Robertson uncertainty relations of $\ob{A}_0 \ob{B}_j$ and $\ob{A}_1
\ob{B}_j$, the inequality \eqref{eq:Sun}.

\subsection{Nonlocality and Heisenberg uncertainty}

An interesting corollary of Theorem~\ref{thm:main1} is that there is a
bound, a generalization of Tsirelson's $2 \sqrt{2}$, for different values of
$\eta_{\ob{A}}$ and $\eta_{\ob{B}}$. In particular,
\begin{equation}
\label{eq:Tsir}
\abs{\mathscr{B}} \leq 2 \sqrt{2} \sqrt{1 - \max\{
  \eta_{\ob{A}}^2, \, \eta_{\ob{B}}^2  \}}
\end{equation}
A geometrical view of this bound is given in Figure~\ref{fig:ind}.
Application of \eqref{eq:Tsir} to $\ob{A}_0 = \ob{x}$ and
$\ob{A}_1 = \ob{p}$, the position and momentum operators, yields
\begin{equation}
\abs{\mathscr{B}} \leq 2 \sqrt{2} \sqrt{1 - \left(\frac{\hbar /
      2}{\cv{\ob{x}} \cv{\ob{p}}}\right)^2}
\end{equation}
which follows from the definition of $\eta_{\ob{A}}$ and the identity
$[\ob{x},\ob{p}] = i \hbar$. This elucidates the interplay between the extent of
nonlocality and the Heisenberg uncertainty principle. The greater the
uncertainty $\cv{\ob{x}} \cv{\ob{p}}$, the stronger the nonlocality
may get, where Tsirelson's $2 \sqrt{2}$ corresponds to the limit
$\cv{\ob{x}} \cv{\ob{p}} \to \infty$.

More generally, relativistic independence
implies a close relationship between nonlocality as quantified by the
Bell-CHSH parameter and the uncertainty parameter $r$ in
\eqref{eq:uncg}. This is summarized in the next theorem.

\begin{theorem}
By relativistic independence
\begin{equation}
\left( \frac{\mathscr{B}}{2 \sqrt{2}}\right)^2 + \abs{r'}^2 \leq 1
\end{equation}
where as before, $r' \ass \frac{r}{\cv{A_1} \cv{A_0}}$. In quantum
mechanics where $r = r_{\text{Q}}$ in \eqref{eq:a} this relation
assumes an explicit form
\begin{equation}
\left( \frac{\mathscr{B}}{2 \sqrt{2}}\right)^2 + \abs{\frac{\langle
    \ob{A}_0 \ob{A}_1 \rangle - \langle \ob{A}_0 \rangle \langle
    \ob{A}_1 \rangle}{\cv{\ob{A}_0} \cv{\ob{A}_1}}}^2 \leq 1
\end{equation}
\end{theorem}

\emph{Proof}. Assume that $\varrho_{ij} = (-1)^{ij} \varrho$, a
configuration underlying the maximal Bell-CHSH parameter, i.e.,
$\mathscr{B} = 4 \varrho$. Relativistic independence \eqref{eq:uncg}
implies \eqref{eq:covineq}, which in this case yields
\begin{equation}
\left[r' - (-1)^j \varrho^2 \right]^2 \leq \left( 1 - \varrho^2 \right)^2
\end{equation}
That is,
\begin{equation}
\label{eq:jj}
\abs{r'}^2 + \left( \frac{\mathscr{B}}{2 \sqrt{2}} \right)^2 -2 (-1)^j r' \varrho^2 \leq 1
\end{equation}
where we have used the identity $\varrho = \mathscr{B} / 4$. Averaging
\eqref{eq:jj} for $j=0$ and $j=1$ implies the theorem.

\subsection{Locality from relativistic independence}

The preceding sections forged a theory-free notion of nonlocality in the form of
correlators that satisfy relativistic independence. Can locality (as appearing in classical statistical theories), which is normally defined by means of Bell inequalities, be similarly
characterized? We will show that locality is in some sense a variant
of relativistic independence.

The first relativistic independence condition in \eqref{eq:uncg} may alternatively be
written as
\begin{equation}
\label{eq:q_cor}
\mathcal{M}^Q \ass
\begin{bmatrix}
M^{-1} \cvm{A} M^{-1} - \tilde{R}_0 \tilde{R}_0^T & 0_{2 \times 2} \\
0_{2 \times 2} & M^{-1} \cvm{A} M^{-1} - \tilde{R}_1 \tilde{R}_1^T
\end{bmatrix} \succeq 0
\end{equation}
where $M$ is a diagonal matrix whose (non-vanishing) entries are
$\cv{A_1}$ and $\cv{A_0}$, and $\tilde{R}_j^T = [\varrho_{0j}, \,
\varrho_{1j}]$. Relativistic independence may further restrict the underlying
correlators when the off-diagonal blocks do not vanish. Locality is implied, for example, by
\begin{equation}
\label{eq:local_cor}
\mathcal{M}^L \ass
\begin{bmatrix}
M^{-1} \cvm{A} M^{-1} - \tilde{R}_0 \tilde{R}_0^T & \tilde{R}_0
\tilde{R}_1^T \\
\tilde{R}_1 \tilde{R}_0^T & M^{-1} \cvm{A} M^{-1} - \tilde{R}_1 \tilde{R}_1^T
\end{bmatrix} \succeq 0
\end{equation}
In particular,
\begin{equation}
u \mathcal{M}^L u^T = 4 - \mathscr{B}^2 \geq 0
\end{equation}
where $u = [1,1,1,-1]$, and $\mathscr{B} \ass \varrho_{00} +
\varrho_{10} + \varrho_{01} - \varrho_{11}$ is the Bell-CHSH
parameter.

 The non-vanishing off-diagonal matrices in
  \eqref{eq:local_cor} essentially render the underlying uncertainty
  relations of both experimenters ineffective. To see how,
  note that the matrix in \eqref{eq:local_cor} (but not that in
  \eqref{eq:q_cor}) is the covariance
  of the four products $A_i B_j$, $i,j = 0,1$, where $A_i$ and $B_j$
  are Alice's and Bob's measurement outcomes. Therefore, the joint
  probabilities of $A_0$ and $A_1$, and of $B_0$ and $B_1$ exist and
  the correlators satisfy the Bell-CHSH inequality. As mentioned in
  the main text, here the parameter $r = \cvv(A_0, A_1)$, and
  $\cv{A_0}^2 \cv{A_1}^2 \geq r^2$. However, this form of the
  uncertainty relation cannot be saturated but for the trivial case of
deterministic $A_0$ and $A_1$.


\subsection{Relativistic independence in general multipartite settings}

Suppose that some experimenters are located at spacetime region $S$
and some others at spacetime region $T$. Each
experimenter has an arbitrary number of measuring devices. We shall
denote by $S_i$ and $T_j$ the vectors of measurements in $S$ and $T$,
where the indices $i$ and $j$ represent sets of choices of measuring
devices in each region. As in the bipartite case, we may write
$\cvm{S}(i)$ and $\cvm{T}(j)$ for the uncertainty relations underlying
the sets of measurements $i$ in $S$ and $j$ in $T$. The covariances
between $S_i$ and $T_j$ may similarly be expressed by a matrix $R$.

Relativistic independence dictates that uncertainty relations in $S$ are
independent of choices in $T$. Therefore, $S$ is
independent of whether $j=0$ or $j=1$ in $T$. This is expressed
mathematically by
\begin{equation}
\label{eq:mult}
\begin{bmatrix}
\cvm{T}(0) & R_0^T \\
R_0 & \cvm{S}
\end{bmatrix} \succeq 0, \quad
\begin{bmatrix}
\cvm{T}(1) & R_1^T \\
R_1 & \cvm{S}
\end{bmatrix} \succeq 0
\end{equation}
But also in the converse direction, uncertainty relations in $T$ are
independent of choices in $S$,
\begin{equation}
\label{eq:mult1}
\begin{bmatrix}
\cvm{T} & \bar{R}_0^T \\
\bar{R}_0 & \cvm{S}(0)
\end{bmatrix} \succeq 0, \quad
\begin{bmatrix}
\cvm{T} & \bar{R}_1^T \\
\bar{R}_1 & \cvm{S}(1)
\end{bmatrix} \succeq 0
\end{equation}
Below we use these to derive a
bound on the quantum mechanical, Alice-Bob, Alice-Charlie, and
Bob-Charlie, one- and two-point correlators. The relation thus
obtained generalizes \eqref{eq:qb} in this tripartite setting.

We note that \eqref{eq:mult} and \eqref{eq:mult1} do not represent the
most general approach for characterizing nonlocal
correlations. Nevertheless, they facilitate analyses and in particular
the derivation of the theorems that follow. A complete
characterization of the set of quantum correlations would require
analyzing \eqref{eq:u} in a general multipartite setting. In such a
case the cross-correlations between the $S$ and $T$ subsets would have
to be accounted for. To some degree this is practiced in the derivation
of Theorem~\ref{thm:trinew1} where it is assumed that Bob and Charlie are
correlated. Disconnecting them by making their correlations zero leads
to the well known monogamy relation in Theorem~\ref{thm:mono}.

In the tripartite case, where Alice in $S$ measures
either $A_0$ or $A_1$, and Bob and Charlie in $T$ measure $(B_l, \,
C_k)$ or $(B_{l'}, \, C_{k'})$, relativistic independence \eqref{eq:mult} holds for
\begin{equation}
\label{eq:trip}
\cvm{T}(0) \ass
\begin{bmatrix}
\cv{C_k}^2 & \cvv(C_k, \, B_l)\\
\cvv(C_k, \, B_l) & \cv{B_l}^2
\end{bmatrix}, \; \;
\cvm{T}(1) \ass
\begin{bmatrix}
\cv{C_{k'}}^2 & \cvv(C_{k'}, \, B_{l'})\\
\cvv(C_{k'}, \, B_{l'}) & \cv{B_{l'}}^2
\end{bmatrix}, \; \;
\cvm{S} \ass
\begin{bmatrix}
\cv{A_1}^2 & r\\
r & \cv{A_0}^2
\end{bmatrix}
\end{equation}
where
\begin{equation}
\label{eq:trip1}
R_0^T =
\begin{bmatrix}
\cvv(A_1, \, C_k) & \cvv(A_0, \, C_k) \\
\cvv(A_1, \, B_l) & \cvv(A_0, \, B_l)
\end{bmatrix}, \; \;
R_1^T =
\begin{bmatrix}
\cvv(A_1, \, C_{k'}) & \cvv(A_0, \, C_{k'}) \\
\cvv(A_1, \, B_{l'}) & \cvv(A_0, \, B_{l'})
\end{bmatrix}
\end{equation}

\begin{theorem}
\label{thm:trinew1}
The relativistic independence condition \eqref{eq:mult} with the matrices in
\eqref{eq:trip} and \eqref{eq:trip1} imply
\begin{equation}
\label{eq:ttbound}
\abs{\zeta_{01}(l,k) - \zeta_{01}(l',k')} \leq \sqrt{(1 -
  \zeta_{11}(l,k)) ( 1 - \zeta_{00}(l,k))} + \sqrt{(1 -
  \zeta_{11}(l',k')) ( 1 - \zeta_{00}(l',k'))}
\end{equation}
where,
\begin{equation}
\zeta_{ij}(l,k) \ass \left[ \varrho^{AC}_{ik} \varrho^{AC}_{jk} -
  \varrho^{BC}_{lk} \varrho^{AB}_{il} \varrho^{AC}_{jk} -
  \varrho^{BC}_{lk} \varrho^{AB}_{jl} \varrho^{AC}_{ik} +
  \varrho^{AB}_{il} \varrho^{AB}_{jl} \right] / \left ( 1 -
  (\varrho^{BC}_{lk})^2 \right)
\end{equation}
and $\varrho^{XY}_{ij} \ass
\cvv(X_i, \, Y_j) / (\cv{X_i} \cv{Y_j})$. Note that letting
$\varrho^{AC} = \varrho^{BC} = 0$ in \eqref{eq:ttbound} recovers the
bound on the Alice-Bob correlators, the first inequality in \eqref{eq:qb}.
\end{theorem}

\emph{Proof}.
Substituting \eqref{eq:trip} into \eqref{eq:mult} yields
\begin{equation}
\small
\cvm{ABC} \ass
\begin{bmatrix}
\cv{C_k}^2 & \cvv(C_k, \, B_l) & \cvv(C_k, \, A_1) & \cvv(C_k, \,
A_0) \\
\cvv(C_k, \, B_l) & \cv{B_l}^2 & \cvv(B_l, \, A_1) & \cvv(B_l, \,
A_0) \\
\cvv(C_k, \, A_1) & \cvv(B_l, \, A_1) & \cv{A_1}^2 & r
\\
\cvv(C_k, \, A_0) & \cvv(B_l, \, A_0) & r & \cv{A_0}^2
\end{bmatrix} \succeq 0
\end{equation}
and similarly for $k'$ and $l'$. This is equivalent to
\begin{equation}
\label{eq:gr1}
M^{-1} \cvm{ABC} M^{-1} =
\begin{bmatrix}
1 & \varrho^{BC}_{lk} & \varrho^{AC}_{1k} & \varrho^{AC}_{0k} \\
\varrho^{BC}_{lk} & 1 & \varrho^{AB}_{1l} & \varrho^{AB}_{0l} \\
\varrho^{AC}_{1k} & \varrho^{AB}_{1l} & 1 & r' \\
\varrho^{AC}_{0k} & \varrho^{AB}_{0l}  & r' & 1
\end{bmatrix} \succeq 0
\end{equation}
where $r' \ass r / (\cv{A_1} \cv{A_0})$, and $M$ is a diagonal matrix
whose (non-vanishing) entries are $\cv{C_k}$, $\cv{B_l}$, $\cv{A_1}$,
and $\cv{A_0}$. By the Schur complement condition for positive
semidefiniteness, \eqref{eq:gr1} is equivalent to
\begin{equation}
\label{eq:gr}
\begin{bmatrix}
1 & r' \\
r' & 1
\end{bmatrix} \succeq
\begin{bmatrix}
\varrho^{AC}_{1k} & \varrho^{AC}_{0k} \\
\varrho^{AB}_{1l} & \varrho^{AB}_{0l}
\end{bmatrix}^T
\begin{bmatrix}
1 & \varrho^{BC}_{lk} \\
\varrho^{BC}_{lk} & 1
\end{bmatrix}^{-1}
\begin{bmatrix}
\varrho^{AC}_{1k} & \varrho^{AC}_{0k} \\
\varrho^{AB}_{1l} & \varrho^{AB}_{0l}
\end{bmatrix}
\end{equation}
which holds if and only if the determinant of the matrix obtained by
subtracting the right-hand side from the left-hand side in \eqref{eq:gr} is nonnegative.
Carrying out this calculation for $k$,$l$ and then for $k'$,$l'$, and
invoking the triangle inequality yield \eqref{eq:ttbound}.

The next theorem shows that the bound \eqref{eq:ttbound} implies
monogamy of correlations. This means that breaking of monogamy
necessarily violates relativistic independence.

\begin{theorem}
\label{thm:mono}
If Charlie and Bob are uncorrelated, $\cvv(C_k, \, B_j) = 0$, then
by relativistic independence
\begin{equation}
\label{eq:mono}
\mathscr{B}_{AB}^2 + \mathscr{B}_{AC}^2 \leq 8
\end{equation}
and therefore also $\abs{\mathscr{B}_{AB}} + \abs{\mathscr{B}_{AC}}
\leq 4$, where both Bell-CHSH parameters, $\mathscr{B}_{AB}$ and
$\mathscr{B}_{AC}$, are for the same pair, $A_0$, $A_1$.
\end{theorem}

\emph{Proof}. Substituting $\varrho^{BC}_{jk} = 0$ in \eqref{eq:gr}
implies
\begin{equation}
\label{eq:se}
2(1 \pm r') = u^T
\begin{bmatrix}
1 & r'\\
r' & 1
\end{bmatrix} u \geq
u^T
\begin{bmatrix}
\varrho^{AC}_{1k} & \varrho^{AB}_{1j} \\
\varrho^{AC}_{0k} & \varrho^{AB}_{0j}
\end{bmatrix}
\begin{bmatrix}
\varrho^{AC}_{1k} & \varrho^{AB}_{1j} \\
\varrho^{AC}_{0k} & \varrho^{AB}_{0j}
\end{bmatrix}^T u =
\left[ \varrho^{AB}_{0j} \pm \varrho^{AB}_{1j} \right]^2 + \left[
  \varrho^{AC}_{0k} \pm \varrho^{AC}_{1k} \right]^2
\end{equation}
for $u^T = [1, \, \pm 1]$. Therefore,
\begin{equation}
4 \geq \left[ \varrho^{AB}_{00} \pm \varrho^{AB}_{10} \right]^2 + \left[
  \varrho^{AC}_{00} \pm \varrho^{AC}_{10} \right]^2 + \left[ \varrho^{AB}_{01} \pm \varrho^{AB}_{11} \right]^2 + \left[
  \varrho^{AC}_{01} \pm \varrho^{AC}_{11} \right]^2 \geq \frac{1}{2}
\mathscr{B}_{AB}^2 + \frac{1}{2} \mathscr{B}_{AC}^2
\end{equation}
from which the theorem follows.

\subsection{Monogamy of correlations in general multipartite
  settings}

The above result is a special case of the more general scenario where
any number of experimenters are correlated with Alice but uncorrelated
among themselves. Suppose there are $n$ experimenters whose
measurements are uncorrelated, $\cvv(M_i^{k}, \, M_j^{l}) = 0$, where
$M_i^{k}$ stands in for the $k$th physical variable measured by the
$i$th experimenter. In this case the generalized uncertainty relations underlying Alice
measurements $A_0, \, A_1$, and the $n$ other measurements $M_1^{i_1},
\ldots, M_n^{i_n}$ are described by
\begin{equation}
\label{eq:multmono}
\left[
\begin{array}{c|cc}
                  & \varrho_{0,i_1}^1 & \varrho_{1,i_1}^1 \\
I_{n \times n} & \vdots & \vdots \\
                  & \varrho_{0,i_n}^n & \varrho_{1,i_n}^n \\
\hline
                  & 1 & r'_{i_1,\ldots,i_n} \\
                  &    &            1
\end{array} \right] \succeq 0
\end{equation}
where, $\varrho_{i, k}^s \ass \cvv(A_i, \, M_s^k) / (\cv{A_i} \cv{M_s^k})$.
This matrix is obtained as an extension of \eqref{eq:u3} following a normalization
similar to the one in previous sections. In this case, Alice's
uncertainty relations are governed by the parameter
$r'_{i_1,\ldots,i_n}$ which may depend on the choices of all of the other
experimenters.

\begin{theorem}
Relativistic independence implies
\begin{equation}
\sum_{s=1}^n \abs{\mathscr{B}_s} \leq \sqrt{2n} \left( \sqrt{1 + r'} +
\sqrt{1 - r'} \right) \leq 2 \sqrt{2n}
\end{equation}
where $\mathscr{B}_s \ass \varrho_{0, i_s}^s + \varrho_{1, i_s}^s +
\varrho_{0, j_s}^s - \varrho_{1,j_s}^s$ is the Bell-CHSH parameter of
Alice and the $s$th experimenter.
Tsirelson's bound and the monogamy property of correlations follow
from this inequality as special cases for $n=1$ and $n=2$,
respectively.
\end{theorem}

\emph{Proof}. If relativistic independence holds then
$r'_{i_1,\ldots,i_n} = r'_{j_1,\ldots,j_n} = r'$. By the Schur
complement condition for positive semidefiniteness,
\eqref{eq:multmono} is equivalent to
\begin{equation}
\label{eq:newmono}
\begin{bmatrix}
1 & r' \\
r' & 1
\end{bmatrix} \succeq
\sum_{s=1}^n \begin{bmatrix}
\varrho_{0,i_s}^s \\
\varrho_{1,i_s}^s
\end{bmatrix}
\begin{bmatrix}
\varrho_{0,i_s}^s &
\varrho_{1,i_s}^s
\end{bmatrix}
\end{equation}
and similarly,
\begin{equation}
\label{eq:newmono1}
\begin{bmatrix}
1 & r' \\
r' & 1
\end{bmatrix} \succeq
\sum_{s=1}^n \begin{bmatrix}
\varrho_{0,j_s}^s \\
\varrho_{1,j_s}^s
\end{bmatrix}
\begin{bmatrix}
\varrho_{0,j_s}^s &
\varrho_{1,j_s}^s
\end{bmatrix}
\end{equation}
Both \eqref{eq:newmono} and \eqref{eq:newmono1} imply
\begin{equation}
2(1 \pm r') \geq \sum_{s=1}^n \left( \varrho_{0,i_s}^s \pm
  \varrho_{1,i_s}^s \right)^2, \quad 2(1 \pm r') \geq \sum_{s=1}^n \left( \varrho_{0,j_s}^s \pm
  \varrho_{1,j_s}^s \right)^2
\end{equation}
which are obtained similarly to \eqref{eq:se}. By norm equivalence,
\begin{equation}
2n(1 \pm r') \geq \left( \sum_{s=1}^n \abs{\varrho_{0,i_s}^s \pm
    \varrho_{1,i_s}^s} \right)^2, \quad 2n(1 \pm r') \geq \left( \sum_{s=1}^n \abs{\varrho_{0,j_s}^s \pm
    \varrho_{1,j_s}^s} \right)^2
\end{equation}
Finally, invoking the triangle inequality
\begin{equation}
\sum_{s=1}^n \abs{\mathscr{B}_s} \leq \sum_{s=1}^n \abs{\varrho_{0,i_s}^s +
    \varrho_{1,i_s}^s} + \abs{\varrho_{0,j_s}^s -
    \varrho_{1,j_s}^s} \leq \sqrt{2n(1 + r')} + \sqrt{2n(1-r')} \leq 2
  \sqrt{2n}
\end{equation}

\subsection{Tighter than Schr\"{o}dinger-Robertson uncertainty
  relations following from \eqref{eq:uncg}}

Alice's uncertainty relations are represented by the $2 \times 2$ lower
submatrix $\cvm{A}$ in the generalized uncertainty relation
\eqref{eq:uncg}. This shows that \eqref{eq:uncg} is more stringent than
any uncertainty relation derived exclusively from $\cvm{A} \succeq 0$. Consider,
for example, a generalized uncertainty relation of the form
\begin{equation}
\label{eq:uuu}
\begin{bmatrix}
\cvm{D} & C \\
C^T & \cvm{A}
\end{bmatrix} \succeq 0
\end{equation}
where $D$ is an invertible $n \times n$ matrix, and $C$ is $n \times
2$ cross-covariance matrix. By the Schur complement condition for
positive semidefiniteness this inequality is equivalent to $\cvm{A}
\succeq C^T \cvm{D}^{-1} C$, which unless $C$ vanishes is tighter than
$\cvm{A} \succeq 0$.

As shown in the preceding sections, from within quantum mechanics the
inequality $\cvm{A} \succeq 0$, which follows from the lower $2 \times
2$ submatrix in \eqref{eq:u3} and \eqref{eq:uncg}, is equivalent to the
Schr\"{o}dinger-Robertson uncertainty relations underlying Alice's
observables $\ob{A}_0$ and $\ob{A}_1$. That quantum mechanics obey
generalized uncertainty relations like \eqref{eq:uncg}, and more
generally \eqref{eq:uuu}, implies that any uncertainty relation
derived from $\cvm{A} \succeq 0$ makes only a small part of the
story. There are many more restrictions arising from our approach all
of which are tighter than the Schr\"{o}dinger-Robertson uncertainty relation that
are obeyed by Alice's observables. One such uncertainty relation is
given below.

Let $D = \ob{A}_i^m$, where $\ob{A}_i$ is one of Alice's observables,
$i=0,1$, and $m$ is an integer, $m > 1$. From within quantum
mechanics, the generalized uncertainty \eqref{eq:uuu} is now given by
\begin{equation}
\label{eq:ng}
\begin{bmatrix}
\cv{\ob{A}_i^m}^2 & \cvv(\ob{A}_i^m, \, \ob{A}_1) & \cvv(\ob{A}_i^m, \,
\ob{A}_0) \\
\cvv(\ob{A}_1, \, \ob{A}_i^m) & \cv{\ob{A}_1}^2 & \cvv(\ob{A}_1, \,
\ob{A}_0) \\
\cvv(\ob{A}_0, \, \ob{A}_i^m) & \cvv(\ob{A}_0, \, \ob{A}_1) &
\cv{\ob{A}_0}^2
\end{bmatrix} \succeq 0
\end{equation}
where $\cvv(\ob{A}_i, \, \ob{A}_j) \ass \langle \ob{A}_i \ob{A}_j
\rangle - \langle \ob{A}_i \rangle \langle \ob{A}_j \rangle$. The
quantities $\cv{\ob{A}_i^m}^2$ and $\cvv(\ob{A}_i^m, \, \ob{A}_1)$ in
\eqref{eq:ng} involve higher statistical moments of the underlying
observables. The inequality \eqref{eq:ng} is equivalent to
\begin{equation}
\cvm{A} =
\begin{bmatrix}
\cv{\ob{A}_1}^2 & \cvv(\ob{A}_1, \,
\ob{A}_0) \\
\cvv(\ob{A}_0, \, \ob{A}_1) &
\cv{\ob{A}_0}^2
\end{bmatrix}
\succeq
\cv{\ob{A}_i^m}^{-2}
\begin{bmatrix}
\cvv(\ob{A}_1, \, \ob{A}_i^m) \\
\cvv(\ob{A}_0, \, \ob{A}_i^m)
\end{bmatrix}
\begin{bmatrix}
\cvv(\ob{A}_i^m, \, \ob{A}_1) &
\cvv(\ob{A}_i^m, \, \ob{A}_0)
\end{bmatrix}
\end{equation}
by the Schur complement condition for positive semidefiniteness. Let
$v^T \ass [1, \pm 1] / \sqrt{2}$ and note that
\begin{equation}
v^T \cvm{A} v =
\frac{1}{2} \cv{\ob{A}_1}^2 + \frac{1}{2} \cv{\ob{A}_0}^2 \pm
\left[ \frac{1}{2} \langle \{ \ob{A}_1, \, \ob{A}_0 \} \rangle -
  \langle \ob{A}_1 \rangle \langle \ob{A}_0 \rangle \right] \geq
\frac{1}{2 \cv{\ob{A}_i^m}^2}
\abs{\cvv(\ob{A}_1, \, \ob{A}_i^m) \pm \cvv(\ob{A}_0, \, \ob{A}_i^m) }^2
\end{equation}
Therefore,
\begin{equation}
\cv{\ob{A}_1}^2 + \cv{\ob{A}_0}^2 \geq
2 \abs{\frac{1}{2} \langle \{ \ob{A}_1, \, \ob{A}_0 \} \rangle -
  \langle \ob{A}_1 \rangle \langle \ob{A}_0 \rangle} + \frac{1}{\cv{\ob{A}_i^m}^2}
\abs{\cvv(\ob{A}_1, \, \ob{A}_i^m) \pm \cvv(\ob{A}_0, \, \ob{A}_i^m) }^2
\end{equation}
This uncertainty relation is to be contrasted with
\begin{equation}
\cv{\ob{A}_1}^2 + \cv{\ob{A}_0}^2 \geq
2 \abs{\frac{1}{2} \langle \{ \ob{A}_1, \, \ob{A}_0 \} \rangle -
  \langle \ob{A}_1 \rangle \langle \ob{A}_0 \rangle}
\end{equation}
which follows from $\cvm{A} \succeq 0$ using similar arguments. Note
also that much like the Maccone-Pati uncertainty relations~\cite{MacPat}, these additive
inequalities do not become trivial in the case where the state
coincides with an eigenvector of one of the observables.


\subsection{The measurability of $r_j$ in a bipartite setting}

In what follows we examine relativistic independence from a different
perspective. As mentioned in the main text, this condition may be
viewed as the requirement that one experimenter's
uncertainty relations are independent of another experimenters'
choices. We claim that if it weren't so, relativistic causality would
have been necessarily violated. Our argument is based on the
measurability of $r_j$ in Alice's $\cvm{A}^j$.

\begin{lemma}
\label{lem}
There exists an $r_{jk}$ which is independent of $j$ and $k$ such that
\eqref{eq:u3} holds with $\cvv(C_k, B_j) = 0$ if and only if the four
intervals $[d_{jk}(-), \, d_{jk}(+)]$, $j,k \in \{0,1\}$, with the
$d_{jk}(-)$ and $d_{jk}(+)$ given below, all intersect.
\begin{equation}
\label{eq:int}
d_{jk}(\pm) \ass \varrho^{AB}_{0j} \varrho^{AB}_{1j} + \varrho^{AC}_{0k}
\varrho^{AC}_{1k} \pm \sqrt{\left[ 1 - (\varrho^{AB}_{0j})^2 -
    (\varrho^{AC}_{0k})^2\right] \left[ 1 - (\varrho^{AB}_{1j})^2 - (\varrho^{AC}_{1k})^2\right]}
\end{equation}
\end{lemma}

\emph{Proof}.
The inequality \eqref{eq:u3} may be written as
\begin{equation}
\label{eq:wnew}
M^{-1} \cvm{ABC}^{jk} M^{-1} =
\begin{bmatrix}
1 & \varrho^{BC}_{jk} & \varrho^{AC}_{1k} & \varrho^{AC}_{0k} \\
\varrho^{BC}_{jk} & 1 & \varrho^{AB}_{1j} & \varrho^{AB}_{0j} \\
\varrho^{AC}_{1k} & \varrho^{AB}_{1j} & 1 & r'_{jk} \\
\varrho^{AC}_{0k} & \varrho^{AB}_{0j} & r'_{jk} & 1
\end{bmatrix} \succeq 0
\end{equation}
where $r'_{jk} \ass r_{jk} / (\cv{A_0} \cv{A_1})$, and $M$ is a
diagonal matrix whose non-vanishing entries are $\cv{C_k}$,
$\cv{B_j}$, $\cv{A_1}$, and $\cv{A_0}$. As $\varrho^{BC}_{jk} =
\cvv(C_k, \, B_j) / (\cv{B_j} \cv{C_k}) = 0$, the Schur complement
condition for positive semidefiniteness implies that \eqref{eq:wnew}
is equivalent to
\begin{equation}
\label{eq:wnew1}
\begin{bmatrix}
1 & r'_{jk} \\
r'_{jk} & 1
\end{bmatrix} -
\begin{bmatrix}
\varrho^{AC}_{1k} & \varrho^{AB}_{1j} \\
\varrho^{AC}_{0k} & \varrho^{AB}_{0j}
\end{bmatrix}
\begin{bmatrix}
\varrho^{AC}_{1k} & \varrho^{AB}_{1j} \\
\varrho^{AC}_{0k} & \varrho^{AB}_{0j}
\end{bmatrix}^T
\succeq 0
\end{equation}
which holds if and only if the diagonal entries obey, $1 -
(\varrho^{AB}_{ij})^2 - (\varrho^{AC}_{ik})^2 \geq 0$, $i=0,1$, and
the determinant of this matrix satisfies
\begin{equation}
\left[ 1 - (\varrho^{AB}_{0j})^2 -
    (\varrho^{AC}_{0k})^2\right] \left[ 1 - (\varrho^{AB}_{1j})^2 -
    (\varrho^{AC}_{1k})^2\right] - \left( r'_{jk} - \varrho^{AB}_{0j} \varrho^{AB}_{1j} - \varrho^{AC}_{0k}
\varrho^{AC}_{1k}\right)^2 \geq 0
\end{equation}
Namely, \eqref{eq:wnew1} holds if and only if
\begin{equation}
\abs{ r'_{jk} - \varrho^{AB}_{0j} \varrho^{AB}_{1j} -
  \varrho^{AC}_{0k} \varrho^{AC}_{1k}} \leq \sqrt{\left[ 1 - (\varrho^{AB}_{0j})^2 -
    (\varrho^{AC}_{0k})^2\right] \left[ 1 - (\varrho^{AB}_{1j})^2 -
    (\varrho^{AC}_{1k})^2\right]}
\end{equation}
for $j,k \in \{0,1\}$. It thus follows that $r'_{jk} \in [d_{jk}(-),
\, d_{jk}(+)]$. If these intervals all intersect then there is $r$ and
$r' \ass r / \cv{A_0} \cv{A_1}$ which are independent of $j,k$ such
that $r'_{jk} = r'$. In particular,
\begin{equation}
\max_{j,k} d_{jk}(-) \leq r' \leq \min_{j,k} d_{jk}(+)
\end{equation}
Conversely, if there is such $r'_{jk} = r'$ then the underlying
intervals necessarily intersect.\\[1ex]

Lemma~\ref{lem} shows that in the absence of Charlie,
$\varrho^{AC}_{ik} = \varrho^{BC}_{jk} = 0$, the parameter $r_j$ in a
bipartite Alice-Bob setting satisfies
\begin{equation}
\label{eq:rjj1}
\varrho_{0j} \varrho_{1j} - \sqrt{(1 - \varrho_{0j}^2) (1 -
  \varrho_{1j}^2) } \leq r_j' \leq \varrho_{0j} \varrho_{1j} +
\sqrt{(1 - \varrho_{0j}^2) (1 - \varrho_{1j}^2) }
\end{equation}
where $\varrho_{ij} = \cvv(A_i, B_j) / (\cv{A_i} \cv{B_j})$, and $r'_j
\ass r_j / (\cv{A_0} \cv{A_1})$.

Let $\mathcal{D}_j$ be the range of admissible $r_j$ in
\eqref{eq:rjj1}. Unless $\mathcal{D}_0 \cap \mathcal{D}_1 \neq
\emptyset$, relativistic independence cannot be satisfied. We shall show that
whenever the two intervals $\mathcal{D}_0$ and $\mathcal{D}_1$ do not
intersect, in which case relativistic independence fails, signaling takes
place. Define
\begin{equation}
\epsilon \ass \min_{w_j \in \mathcal{D}_j} \abs{w_0 - w_1}
\end{equation}
It can be recognized that this $\epsilon$ is the smallest of the four
possible numbers
\begin{equation}
\epsilon = \abs{ \varrho_{00} \varrho_{10} - \varrho_{01} \varrho_{11}
  \pm \sqrt{(1 - \varrho_{00}^2) (1 - \varrho_{10}^2) } \pm \sqrt{(1 -
    \varrho_{01}^2) (1 - \varrho_{11}^2) }}
\end{equation}

Assume now that the intervals $\mathcal{D}_0$ and $\mathcal{D}_1$ do not
intersect and thus $\epsilon > 0$. Here is a procedure that Alice may
in principle follow for detecting a signal from Bob using her local
measurements. Let $\pp$ be a set of local parameters describing
Alice's non-trivial system (for practical reasons $\pp$ can be discretized). The
precision is represented for any physical variable $A$ by the variance
$\cv{A}^2(\pp)$. This $\cv{A}^2(\pp)$ can be evaluated empirically by
measuring $A$ in many trials of an experiment while reproducing time
and again the same set $\pp$.

For any real parameter $\theta \in [-\pi,\pi]$, Alice is able to evaluate
\begin{equation}
g(\theta, \pp) \ass \cos(\theta)^2 \frac{\cv{A_0}(\pp)}{\cv{A_1}(\pp)}
+ \sin(\theta)^2 \frac{\cv{A_1}(\pp)}{\cv{A_0}(\pp)}
\end{equation}
Her uncertainty relation \eqref{eq:u} dictates that this quantity is
bounded from below
\begin{equation}
\label{eq:lb}
\min_{\pp} g(\theta, \pp) \geq \max\{ 0, \; r_j' \sin(2\theta) \}
\end{equation}
which follows from $[\cos \theta, \, -\sin \theta] \cvm{A}^j [\cos \theta,
\, -\sin \theta]^T \geq 0$. That Alice
may reach $r_j'$ means that for some $\theta$ there exists a subset of
parameters $\pp^{\star}$ saturating \eqref{eq:lb},
\begin{equation}
\label{eq:lb1}
\min_{\pp} g(\theta, \pp)  = g(\theta,
\pp^{\star}) = r_j' \sin(2\theta)
\end{equation}
which also implies that $\cvm{A}^j$ is a singular matrix and therefore
$\cv{A_0}^2(\pp^{\star}) \cv{A_1}^2(\pp^{\star}) = r_j^2$.


Suppose that Alice and Bob agree in advance to repeat the underlying
experiment $N$ times, for a sufficiently large $N$. Alice may choose
a new set $\pp$ and a device with which to measure in the beginning of
each trail. All this time Bob uses only one of his devices, say the
$j$-th one. Using the measurement outcomes from all these trails,
Alice may approximate $g(\theta, \pp)$ for each $\pp$ in the domain of these
parameters. According to \eqref{eq:lb1} Alice may then evaluate
$\tilde{r}_j'$, an estimate of $r_j'$, using the approximated minimum
of $g(\theta, \pp)$. In practice, her estimate is accurate up to an
error term, $\delta_j$ of the order $\mathcal{O}(1 / \sqrt{N})$,
i.e., $\tilde{r}_j' = r_j' + \delta_j$. It now follows that for
sufficiently large $N$,
\begin{equation}
\abs{\tilde{r}_0' - \tilde{r}_1'} = \abs{r_0' - r_1' + \delta_0 -
  \delta_1} \geq \abs{\epsilon + \mathcal{O}(1 / \sqrt{N})}
\end{equation}
Alice may therefore be able to evaluate a number whose magnitude is as
large as $\epsilon$ and whose sign tells whether Bob measured first
using $j=0$ and then using $j=1$ or the opposite. Of course, if
independence holds, in which case $\epsilon=0$, Alice will not detect
any signal from Bob via her local uncertainty relations.

\newpage

\section*{Acknowledgements}

We are grateful to Nicolas Gisin and Avshalom Elitzur  for helpful discussions and comments that improved the overall presentation of the idea. We
wish especially to express our gratitude to Yakir Aharonov for many insightful
discussions. In addition, we wish to thank to anonymous reviewers for very helpful comments and suggestions. {\bf Funding:} A.C. acknowledges support from Israel Science Foundation Grant No. 1723/16. E.C. acknowledges support form the Engineering Faculty in Bar Ilan University. {\bf Author contributions:} Both authors developed the concepts and
worked out the mathematical proofs. {\bf Competing interests:} The authors declare that they have no competing
interests. {\bf Data and materials availability:} All data needed to evaluate the conclusions in the paper are present in the paper and/or the Supplementary Materials. Additional data related to this paper may be requested from the authors.

\end{document}